\RequirePackage{ifpdf}
\documentclass[letterpaper]{JHEP3}
\usepackage{amsmath}
\usepackage{epsfig}
\usepackage{subfigure}

\newcommand{\roughly}[1]{\mathrel{\raise.3ex\hbox{$#1$\kern-0.85em
\lower1ex\hbox{$\sim$}}}}

\newcommand{\lsim}{\roughly<}
\newcommand{\gsim}{\roughly>}
\def\pd{\partial}

\def\cD{{\cal D}}
\def\cC{{\cal C}}

\def\cG{{\cal G}}

\def\cK{{\cal K}}
\def\cL{{\cal L}}

\def\cO{{\cal O}}
\def\cR{{\cal R}}

\def\cP{{\cal P}}

\def\exd{{\hbox{d}}}

\def\sech{{\rm sech}\,}

\def\bea{\begin{eqnarray}}
\def\eea{\end{eqnarray}}
\def\be{\begin{equation}}
\def\ee{\end{equation}}

\def\ssB{{\scriptscriptstyle A}}
\def\ssB{{\scriptscriptstyle B}}
\def\ssC{{\scriptscriptstyle C}}
\def\ssF{{\scriptscriptstyle F}}

\def\ssD{{\scriptscriptstyle D}}
\def\ssM{{\scriptscriptstyle M}}
\def\ssA{{\scriptscriptstyle A}}
\def\ssN{{\scriptscriptstyle N}}

\def\ssS{{\scriptscriptstyle S}}

\def\O{\mathcal{O}}

\def\nn{\nonumber}

\def\({\left(}
\def\){\right)}

\def\pref#1{(\ref{#1})}

\def\ol{\overline}
\def\vo{\mathcal{V}}

\title{Inflating with Large Effective Fields}

\author{C.P.~Burgess,${}^{1,2,3}$ M.~Cicoli,${}^{4,5,6}$ F.~Quevedo${}^{6,7}$ and M.~Williams${}^{2,3}$\\

${}^1$ PH\,-TH Division, CERN, CH-1211, Gen\`eve 23, Suisse.\\
${}^2$ Department of Physics \& Astronomy, McMaster University\\ \qquad 1280 Main Street West, Hamilton ON, Canada.\\
${}^3$ Perimeter Institute for Theoretical Physics\\
\qquad 31 Caroline Street North, Waterloo ON, Canada.\\
${}^4$ Dipartimento di Fisica e Astronomia, Universit\`a di Bologna,\\ $\qquad$ via Irnerio 46, 40126 Bologna, Italy.\\
${}^5$ INFN, Sezione di Bologna, Italy.\\
${}^6$ Abdus Salam ICTP, Strada Costiera 11, Trieste 34014, Italy.\\
${}^7$ DAMTP, University of Cambridge, Wilberforce Road,  Cambridge, CB3 0WA, UK.
}

\preprint{DAMTP-2014-28\\CERN-PH-TH-2014-072}

\date{\today}

\abstract {We re-examine large scalar fields within effective field theory, in particular focussing on the issues raised by their use in inflationary models (as suggested by BICEP2 to obtain primordial tensor modes). We argue that when the large-field and low-energy regimes coincide the scalar dynamics is most effectively described in terms of an asymptotic large-field expansion whose form can be dictated by approximate symmetries, which also help control the size of quantum corrections. We discuss several possible symmetries that can achieve this, including pseudo-Goldstone inflatons characterized by a coset $G/H$ (based on abelian and non-abelian, compact and non-compact symmetries), as well as symmetries that are intrinsically higher dimensional. Besides the usual trigonometric potentials of Natural Inflation we also find in this way simple {\em large-field} power laws (like $V \propto \phi^2$) and exponential potentials, $V(\phi) = \sum_{k} V_k \; e^{-k \phi/M}$. Both of these can describe the data well and give slow-roll inflation for large fields without the need for a precise balancing of terms in the potential. The exponential potentials achieve large $r$ through the limit $|\eta| \ll \epsilon$ and so predict $r \simeq \frac83(1-n_s)$; consequently $n_s \simeq 0.96$ gives $r \simeq 0.11$ but not much larger (and so could be ruled out as measurements on $r$ and $n_s$ improve). We examine the naturalness issues for these models and give simple examples where symmetries protect these forms, using both pseudo-Goldstone inflatons (with non-abelian non-compact shift symmetries following familiar techniques from chiral perturbation theory) and extra-dimensional models.}

\begin{document}

\section{Introduction}

The very early universe seems to have been a remarkably simple place: all we know --- and we now know a fair amount --- about the properties of primordial fluctuations is consistent with the predictions of the simplest single-field inflationary models \cite{Inflation}. Although there are an impressive number of single-field inflationary models \cite{Single-field-Zoo}, an even more impressively large number of them lay bleeding on the ground after the recent discovery of primordial gravitational waves \cite{BICEP2}.

Although BICEP2 finds the tensor-to-scalar ratio is $r = 0.2^{+ 0.07}_{- 0.05}$, it is likely that better modeling of foregrounds will reduce this value (for instance the preliminary analysis for one such model gives $r = 0.16^{+0.06}_{-0.05}$ \cite{BICEP2}). To be conservative, for the purposes of this paper we simply take the BICEP2 observations to imply
\be \label{BICEP2r}
  r_{\rm exp} \gsim 0.1 \,.
\ee

What makes this interesting is that a great many models do not give $r$ this large, and so would be decisively ruled out if the observation is confirmed. Among other things this includes the majority of (but not all) string-inflation models \cite{StrInfPostPlanck}. One way to see why large $r$ is such a challenge is the Lyth bound \cite{LythBound} that relates a value this large for $r$ to a trans-Planckian range through which the inflaton rolls in simple single-field models; something which many of the known models do not do. Although efforts have been made towards evading the Lyth bound \cite{LythUnbound}, we believe that it should be embraced: if large field displacements are difficult (but not impossible) to obtain then the fact that Nature seems to want them is likely to be very informative.

Within simple inflationary models primordial perturbations ultimately arise as quantum fluctuations in the gravitational field. Are large field excursions consistent with the validity of the semiclassical approximations on which controlled quantum gravity calculations rely? In principle they need not be inconsistent: in the end the semiclassical approximation relies on the low-energy approximation \cite{EFTGrav}, but it is not necessary that large field values must cost a large energy density. Flat directions in supersymmetric field theories --- for which large fields cost precisely zero energy --- provide perhaps the most direct existence proof that large fields and large energies need not be linked.

What is required for a controlled calculation is an understanding of the behaviour of the lagrangian in the large-field regime, including the behaviour of the scalar potential and the inflaton's target-space geometry. If the scalar potential is bounded for large fields, what is its asymptotic behaviour? And what is the large-field limit of the target-space metric? Are these asymptotic forms stable under quantum corrections?

Symmetries and approximate symmetries can help with all of these issues. For instance, if the inflaton is a pseudo-Goldstone boson for an approximate symmetry \cite{pGB}, then in the symmetry limit it becomes a bona-fide Goldstone boson on which the scalar potential cannot depend. Furthermore, for a symmetry breaking pattern where $G$ breaks to a subgroup $H$ the Goldstone fields parameterize the coset space $G/H$ \cite{CCWZ}, on which the $G$-invariant target-space metrics can also be systematically identified, allowing a coherent picture of the full target-space geometry in both the large- and small-field regimes.

These features are not changed appreciably if $G$ is only an approximate symmetry, since by assumption the geometry of the target space is only slightly perturbed. Furthermore, this approximate symmetry can also ensure the whole picture survives quantum effects. But since the scalar potential itself is a symmetry-breaking effect, any regime where it grows without bound (such as for large fields) introduces a worry about the validity of the entire approximate-symmetry picture. Natural inflation \cite{NatInf} --- with a trigonometric potential naturally arising from the weak breaking of an abelian shift symmetry: $\phi \to \phi + \omega \, f$ --- provides the simplest example along these lines (for which the potential is everywhere bounded), but does not exhaust the possibilities.

In this paper we describe a broader class of potentials that are similarly protected by (generalized) shift symmetries, and so for which the inflaton is a pseudo-Goldstone boson. In particular we display simple examples that illustrate the following two more general points:
\begin{itemize}
\item They show that pseudo-Goldstone bosons can enjoy more complicated potentials than the simple trigonometric potential of Natural Inflation. In particular we show that exponential potentials generically arise (all the while keeping a positive definite target-space metric) when non-compact symmetries are considered. (Ours is not an exhaustive study, and more general kinds of potentials than trigonometric or exponential are also likely possible.)
\item Because the geometry of one dimension is not that interesting, the constraints imposed by symmetry on the large-field target-space geometry are most informative when there is more than one pseudo-Goldstone boson. In this case symmetry arguments can dictate the target-space geometry, showing how field redefinitions can be used to relate the large- and small-field regimes (trading complications in the potential for complications in the scalar kinetic terms). We work through the illustrative example of two Goldstone bosons for which the target space is the coset $SU(1,1)/U(1)$. Of course PLANCK and BICEP2 are also informative here since they also constrain the existence of other light fields during inflation through their contributions to isocurvature perturbations \cite{QCDAxion}.
\end{itemize}

In general a proper understanding of the target space geometry and the nature of the small symmetry breaking terms allows an understanding of what kinds of asymptotic forms are appropriate to the scalar potential in the large-field regime.\footnote{The usual formulation of large-field inflation often is cast using specific coordinates in field space, obscuring the freedom to perform field redefinitions that map large fields to small. Appendix \ref{app:fieldredef} briefly reviews how to recast inflationary slow-roll conditions in a more covariant way.} In particular it asks whether or not the potential diverges in the large-field limit (as is assumed when parameterizing it as a positive power of $\phi$), and if it does how quickly it does. For instance, in the $SU(1,1)/U(1)$ example we find that the potential might diverge or be bounded, but in either case naturally admits an expansion in powers of exponentials, $e^{-k \phi/M_p}$, for large $\phi$.

The rest of the paper is set out as follows. Next, in \S\ref{sec:phenomenology}, we review the inflationary phenomenology of single-field models built on the simplest forms for large-field potentials: those that look like $V \propto \phi^p$; $V \sim V_0 - V_1/\phi^q$ and $V \sim V_0 - V_1 \, e^{-k\phi}$. We show how all of these can successfully describe the combined observations (including $r \gsim 0.1$) and identify which parts of parameter space can do so. Then \S\ref{sec:naturalness} identifies the naturalness issues these models face, and show in particular two different ways that symmetries --- 4D generalized shift symmetries, very similar to Natural Inflation, as well as extra-dimensional symmetries --- can generate exponential potentials in the large-field limit. \S\ref{sec:naturalness} also briefly describes some of the proposed UV completions for these models and comments on the parameter ranges that are found in explicit examples. Finally, our conclusions are summarized in \S\ref{sec:conclusions}.

We collect useful material in three appendices. In Appendix A we present a covariant formulation of slow-roll conditions including an invariant definition of large field for  multi-field models. Appendix B presents explicit discussion of simple compact and non-compact cosets $SO(3)/SO(2)$ and $SO(2,1)/SO(2)$ including the construction of invariant metrics and explicit scalar potentials (exponentials and power-law) following standard techniques of chiral perturbation theory. Appendix C is a discussion of the supersymmetric $SU(1,1)/U(1)$ coset including a discussion of $D$ and $F$ terms.

\section{Inflaton phenomenology}
\label{sec:phenomenology}

We start with a short review of the inflationary phenomenology of several classes of single-field models using scalar potentials that plausibly could arise for large fields. Our logic is to take $n_s \simeq 0.96$ to be a known quantity, from which predictions for $r$ are to be explored as a function of the parameters relevant to the potentials of interest.\footnote{More complicated models for which $n_s$ is much closer to (or greater than) 1 are also possible (and well-motivated \cite{MoreComplicated}) provided they also predict more species of light particles, $\Delta N_{\rm eff} \ne 0$ \cite{rNeff}, but we do not pursue these more complicated options here.}

Our goal is simply to identify what kinds of parameters would be required in each case to account for the Planck and BICEP2 data, so we do not (yet) worry here whether or not the potential of interest can be obtained in a natural way from a sensible microscopic UV completion (we have more to say about this in \S\ref{sec:naturalness}, below). We consider in turn power-law potentials and exponential potentials, and find both admit regions of parameter space that can accommodate the BICEP2 data.

In all cases it is assumed that the kinetic energy of the inflaton is canonical (see however Appendix \ref{app:fieldredef}):
\be
 \cL_{\rm kin} = - \frac12 \, \sqrt{-g} \; (\partial \phi)^2 \,.
\ee
Because of this assumption observables may be computed in terms of the slow-roll parameters
in the usual way for single-field models \cite{LyddleRef}
\be \label{eq:nsrinslowroll}
 n_s -1 \simeq -6\epsilon + 2 \eta\,, \qquad
 r \simeq 16 \, \epsilon \,, \qquad
 n_t \simeq -2 \epsilon \,,
\ee
(including, in particular, the standard single-field consistency condition $n_t \simeq - r/8$), where
\be \label{eq:epsetadef}
 \epsilon = \frac12 \left( \frac{M_p V'}{V} \right)^2   \quad \hbox{and} \quad
 \eta = \frac{M_p^2 V''}{V}  \,.
\ee

\subsection{Power-law potentials}

We consider first power-law potentials of the form
\be \label{eq:powerV}
 V(\phi) = V_0 + V_1 \left( \frac{\phi}{M_p} \right)^p \,,
\ee
where for now $p$ can be positive or negative. This is to be regarded as an asymptotic expansion for {\em large} fields rather than small fields, so we see that $V \to \infty$ as $\phi \to \infty$ if $p > 0$, while $V$ remains bounded as $\phi \to \infty$ when $p < 0$.

When $p > 0$ (or $p < 0$) it is natural to take $V_1 > 0$ (or $V_1 < 0$) so that $\phi$ evolves to smaller values during inflation, with inflation ending once the large-field form, eq.~\pref{eq:powerV}, breaks down. (For completeness we also entertain the other sign for $V_1$, in which case a full model would require a second `waterfall' field to become unstable and trigger inflation's end, {\em \`a la} hybrid inflation \cite{Hybrid}.)

\subsubsection*{When $V_0$ is negligible}

In the simplest case $V_0$ is negligible for large $\phi$ and we may work with only the second term of eq.~\pref{eq:powerV}. This is a natural assumption when $p > 0$ due to the growth of the power-law term.

In this case the important slow-roll parameters are
\be
 \epsilon \simeq \frac12 \left( \frac{p M_p}{\phi} \right)^2 \qquad \hbox{and} \qquad
 \eta \simeq p(p-1) \left( \frac{M_p}{\phi} \right)^2  \,,
\ee
which imply that slow roll is always ensured provided only that $\phi$ is large enough relative to $M_p$, without the need to finely adjust the parameters $V_1$ and $p$ appearing in the potential. Since $V$ grows with $\phi$ eventually the low-energy approximation fails, but this can happen at fields larger than the inflationary regime if $V_1$ is small enough: $V \ll M_p^4$ requires $V_1/M_p^4 \ll (\sqrt{2\epsilon}/p)^p$.

For power-law potentials of this type the slow-roll parameters are related by
\be \label{eq:epsvsetaPLV0zero}
 \epsilon \simeq \frac{p \, \eta}{2(p-1)}  \,,
\ee
and so
\be \label{eq:nsrinexpinfl}
 n_s - 1 \simeq - \left( \frac{p+2}{p-1} \right) \eta \qquad \hbox{and} \qquad
 r \simeq \frac{8 p \,\eta}{p-1} = \frac{8 p }{p+2} (1 - n_s) \,.
\ee
For instance, using $n_s \simeq 0.96$ and $p = 2$ gives $r \simeq 0.16$, likely to be in good agreement with the BICEP2 measurement once foregrounds are properly dealt with. Indeed it is hard to be simpler than the case $p=2$, making it a benchmark model against which others are compared \cite{paolo}.

\subsubsection*{General $V_0$}

More generally, if we drop the assumption that $V_0$ is negligible --- such as would be natural for large $\phi$ when $p = -q < 0$ --- we instead find the following slow-roll parameters
\be
 \epsilon \simeq \frac1{2D^2} \left( \frac{p \, V_1}{V_0} \right)^2  \left( \frac{\phi}{M_p} \right)^{2p-2} \quad \hbox{and} \quad
 \eta \simeq \frac{p(p-1)}{D} \left( \frac{V_1}{V_0} \right) \left( \frac{\phi}{M_p} \right)^{p-2} \,,
\ee
where $D = 1 + (V_1/V_0)(\phi/M_p)^p$.

\FIGURE[h!]{
  \centering
\begin{tabular}{c}
	\includegraphics[width=0.6\textwidth]{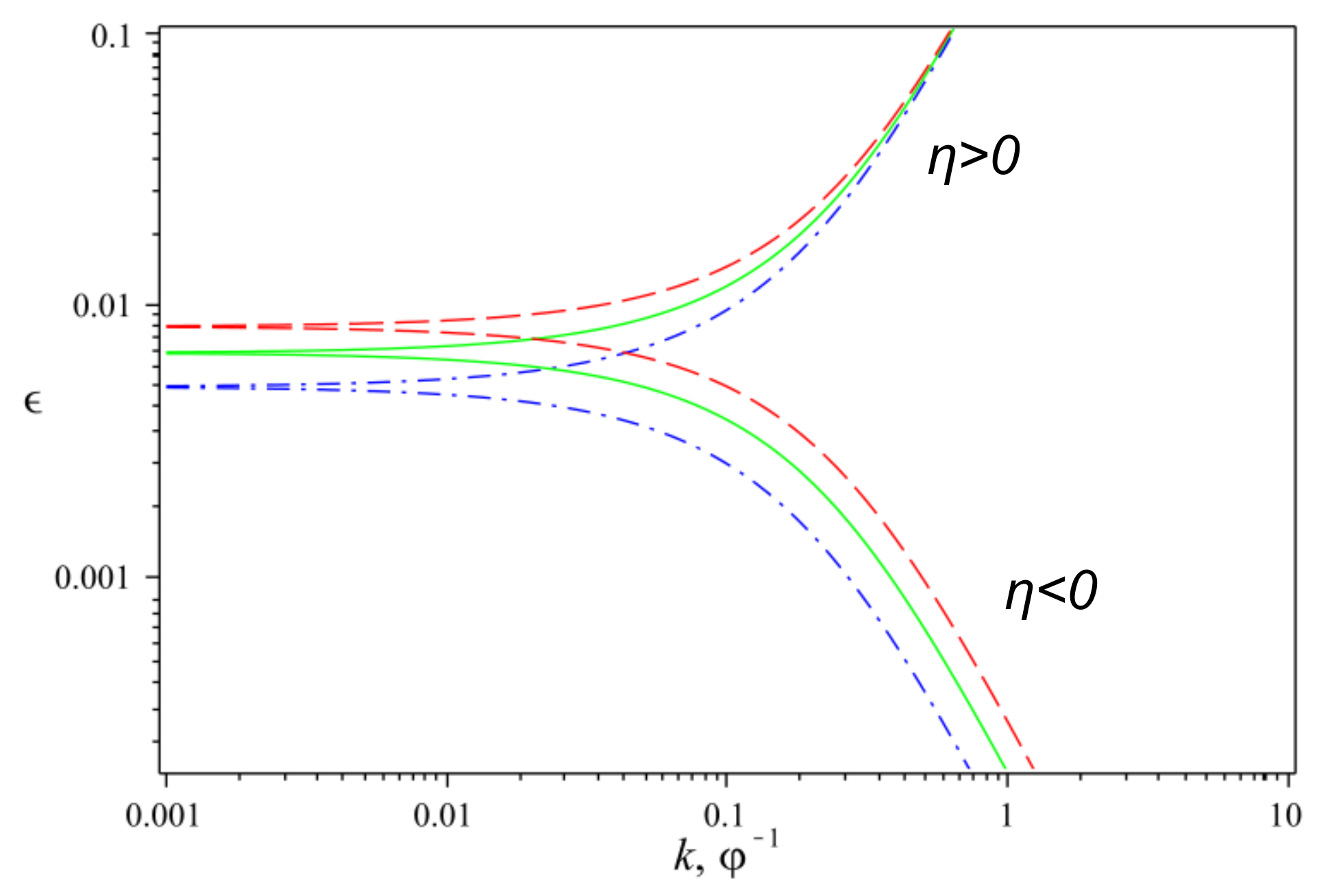} \\
	\includegraphics[width=0.61\textwidth]{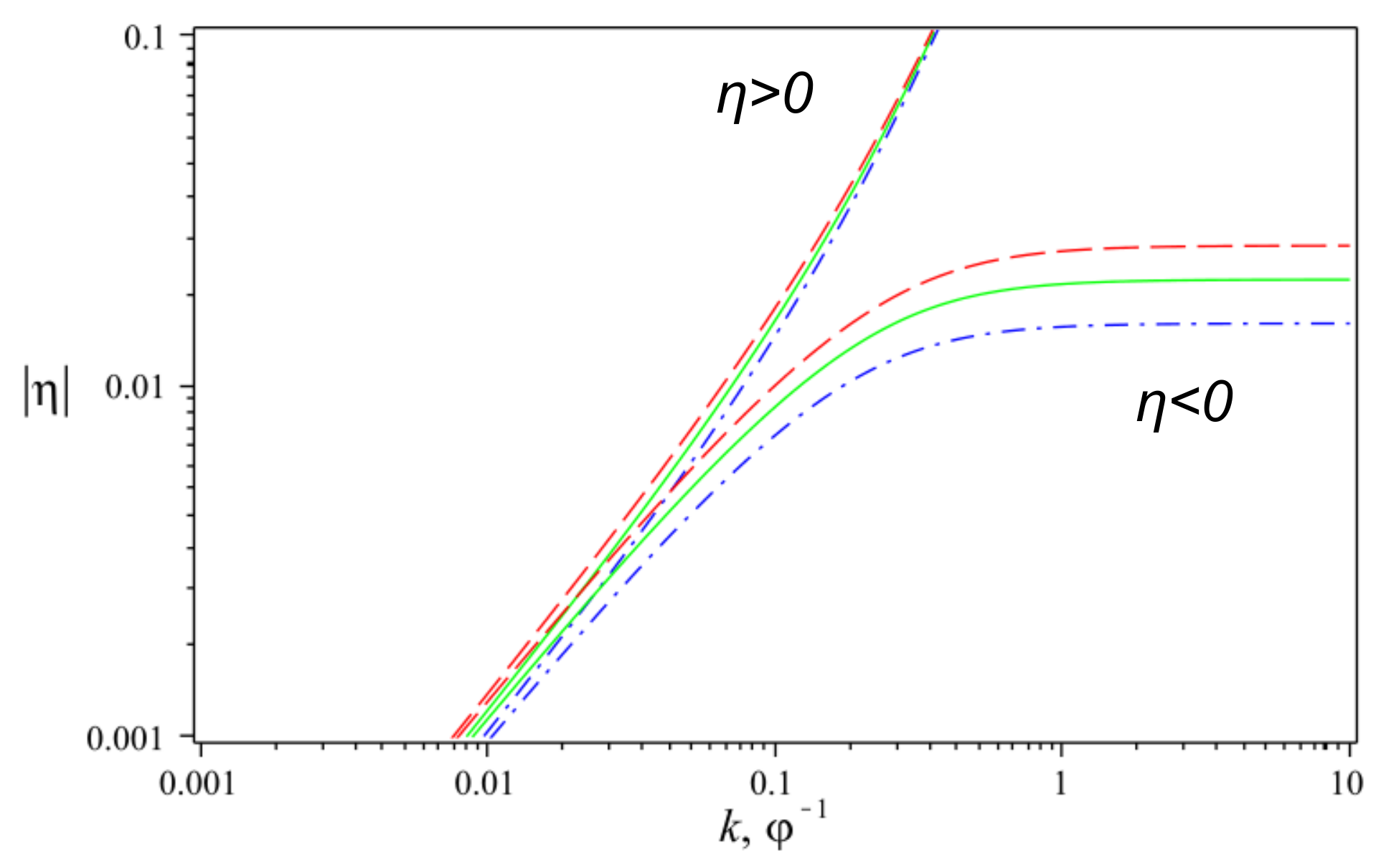}
\end{tabular}
    \caption{Plots of $\epsilon$ and $|\eta|$ for intermediate values of $k$ for exponential potentials (or $\varphi^{-1}$ for power-law potentials). Red dashes and blue dash-dots denote $n_s=0.95$ and $n_s=0.97$, respectively; the green solid line corresponds to $n_s=0.96$. The upper branch applies when $\eta=\eta_+>0$; the lower branch applies when $\eta=\eta_-<0$.}
\label{epsetaplot}
}

Notice that these expressions imply that $\epsilon$ and $\eta$ are related by
\be \label{eq:epsvseta}
 \epsilon \simeq \frac{\varphi^2 \eta^2}{2} \,, \qquad \hbox{where} \qquad
 \varphi := \frac{1}{|p-1|} \left( \frac{\phi}{M_p} \right) \,.
\ee
Using this to eliminate $\epsilon$ from the prediction for $n_s$ we find
\be \label{eq:nsvsvarphip}
 n_s - 1 \simeq - 3 \varphi^2 \eta^2 + 2 \eta \quad \hbox{and} \quad
 r \simeq 8 \,\varphi^2 \, \eta^2 \,.
\ee
Solving eq.~\pref{eq:nsvsvarphip} for $\eta$ we find the two roots
\be \label{eq:PLetaroots}
 \eta_\pm = \frac{1}{3\varphi^2} \Bigl[ 1 \pm \sqrt{ 1 + 3 \varphi^2 (1-n_s)} \Bigr] \,.
\ee
These roots are plotted for $\epsilon$ and $\eta$ in Fig.~\ref{epsetaplot}, which shows that slow roll is generic (and becomes $\varphi$-independent) for large $\varphi$, in which limit $|\eta| \ll \epsilon$. Analytic expressions for this asymptotic behaviour in this limit are easily found by using the large-field limit, $3 \varphi^2 |1 - n_s| \gg 1$ ($\varphi^{-1} \ll 0.35$ when $n_s \simeq 0.96$), in which case the roots become
\be \label{eq:PLetalargephi}
 \eta_\pm \simeq \pm \sqrt{\frac{1 - n_s}{3 \varphi^2}} \qquad
 \hbox{and} \qquad \epsilon_\pm := \frac12 \varphi^2 \eta_\pm^2 \simeq \frac{1 - n_s}{6} \simeq 0.0067 \,,
\ee
where the last equality uses $n_s \simeq 0.96$.

\FIGURE[h!]{
  \centering
  \includegraphics[width=0.7\textwidth]{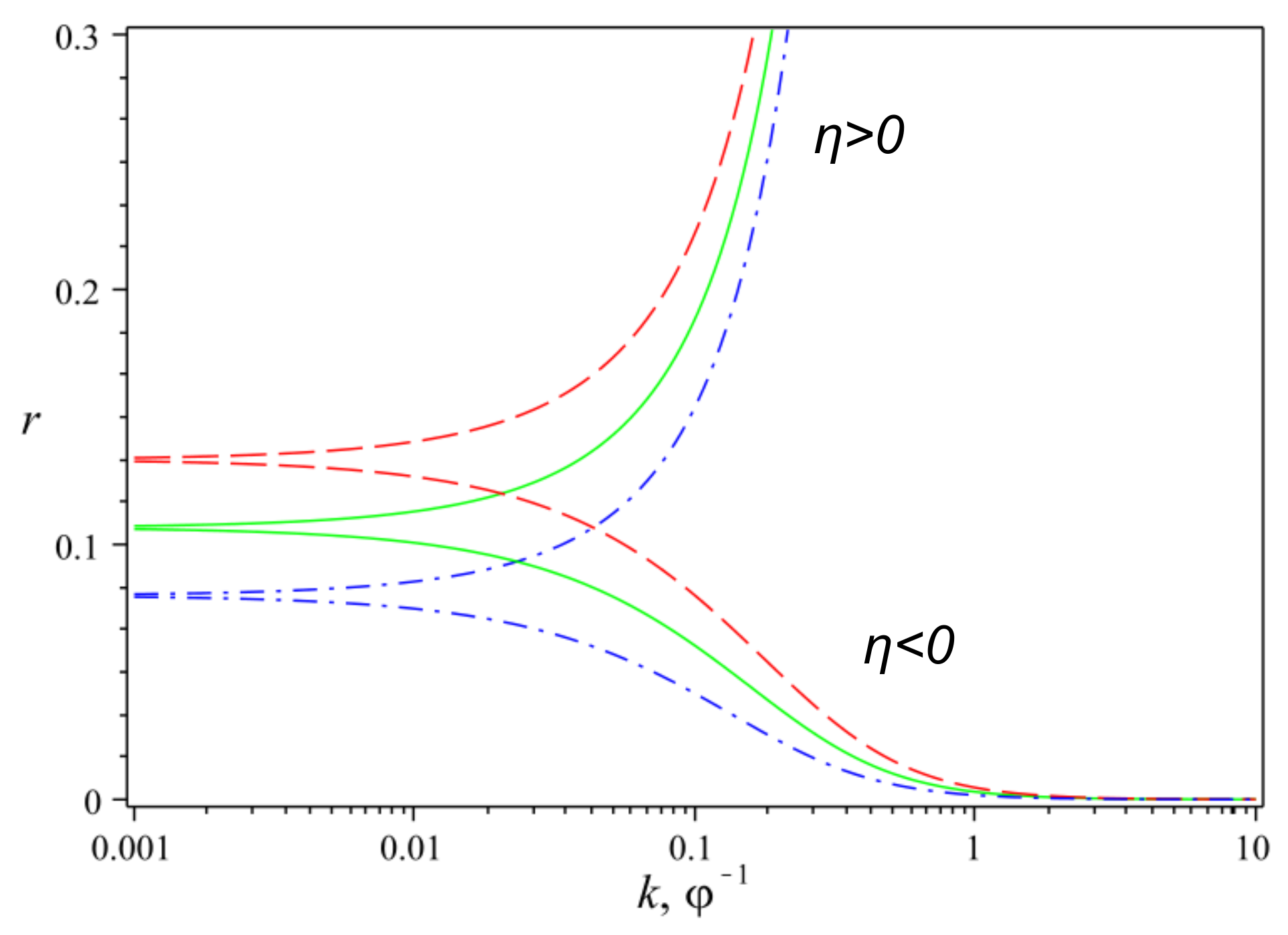}
  \caption{Plot of $r$ for intermediate values of $k$ for exponential potentials (or $\varphi^{-1}$ for power-law potentials). Red dashes and blue dash-dots denote $n_s=0.95$ and $n_s=0.97$, respectively; the green solid line corresponds to $n_s=0.96$. The upper branch applies when $\eta=\eta_+>0$; the lower branch applies when $\eta=\eta_-<0$.}
\label{rplot}
}

The corresponding prediction for $r$ as a function of $\varphi$ obtained from these slow-roll parameters is given in Fig.~\ref{rplot}. In the large-$\varphi$ regime the analytic prediction for $r$ also becomes $\phi$-independent, with
\be \label{eq:runiversal}
 r \simeq \frac83 \, (1 - n_s) \simeq 0.11 \,.
\ee
Again the final equality uses $n_s \simeq 0.96$. This prediction for $r$ is robust inasmuch as it is largely $\varphi$-independent for large $\varphi$, which arises because in this regime $|\eta| \ll \epsilon$. It should be noted, however, that this prediction does {\em not} apply in the simple large-field regime with bounded potential, for which $p = -q < 0$ and $V_0 \gg V_1(M_p/\phi)^q$. This can be seen because in this regime
\be
 \frac{\epsilon}{|\eta|} \ll \frac{p}{2(p-1)} = \frac{q}{2(q+1)} \le \frac12 \quad \hbox{for all} \quad q \ge 0 \,.
\ee

As Fig.~\ref{rplot} also shows, the small-$\phi$ regime also turns out not to describe well the observations. For instance, in the asymptotic regime where $3\varphi^2(1-n_s) \ll 1$ the positive root, $\eta_+$, predicts a value of $r$ that is too large:
\be
 r_+ = 16 \epsilon_+ \simeq \frac{32}{9 \varphi^2} + \frac{16}{3} (1 - n_s) + \cdots \gg 16(1-n_s) \simeq 0.64 \,,
\ee
while, on the other hand, the negative root, $\eta_-$, instead gives a value of $r$ that is much too small to accommodate the BICEP2 data.
\be
 r_- = 16 \epsilon_- \simeq 2 \varphi^2 (1-n_s)^2 \ll \frac23 (1-n_s) \simeq 0.027 \,.
\ee

\subsection{Exponential potentials}

Single-field exponential inflation models instead assume an inflaton potential of the form
\be \label{eq:expV}
 V(\phi) \simeq V_0 - V_1 \; e^{-k \phi/M_p} + \cdots\,,
\ee
where the ellipses denote higher powers of the small quantity $e^{-\phi/M_p}$. Usually we take $k > 0$, but in principle the leading term could be a positive power of $e^{\phi/M_p}$ for large positive $\phi$ provided that the potential does not become too large in the inflationary regime, and provided the subsequent terms in the series are suppressed by powers of $e^{-\phi/M_p}$. If $V_1$ is positive then the roll is towards smaller $\phi$ for which the asymptotic form, eq.~\pref{eq:expV}, eventually fails, ending the slow roll. Occasionally we also entertain negative $V_1$, in which case another field must be invoked to end the inflationary regime.

For this potential the important slow-roll parameters are
\be
 \epsilon = \frac12 \left( \frac{M_p V'}{V} \right)^2 \simeq \frac12 \left( \frac{k \, V_1}{V_0 \cD} \right)^2  e^{-2k \phi/M_p}  \quad \hbox{and} \quad
 \eta = \frac{M_p^2 V''}{V} \simeq - \left( \frac{k^2 V_1}{V_0 \cD} \right) e^{-k \phi/M_p} \,,
\ee
where $\cD = 1 - (V_1/V_0) e^{-k\phi/M_p}$. Again these imply that slow roll is always ensured provided only that $\phi$ is large enough, without the need to finely adjust the parameters $V_0$, $V_1$ and $k$ of the potential. Exponential potentials enjoy the general relation \cite{Fibre}
\be \label{eq:epsvseta}
 \epsilon \simeq \frac{\eta^2}{2k^2} \,,
\ee
and so
\be \label{eq:nsrinexpinfl}
 n_s - 1 \simeq - \frac{3 \eta^2}{k^2} + 2 \eta \quad \hbox{and} \quad
 r \simeq \frac{8 \eta^2}{k^2} \,.
\ee

Solving for $\eta$ in terms of $n_s$ gives precisely the same expressions as found above, eq.~\pref{eq:PLetaroots}, with the replacement $\varphi \to 1/k$. Comparing with the figures then shows that successful models require $k$ large and in the successful regime we have $|\eta| \ll \epsilon$ and so the robust prediction, eq.~\pref{eq:runiversal}, also applies in this regime.

It is again the BICEP2 data that is responsible for the failure of the large-$k$ models. This is easy to see analytically since $k \gsim \O(1)$ ensures $\epsilon \ll |\eta|$ (in the slow-roll regime where $|\eta| \ll 1$) and so $n_s - 1 \simeq 2 \eta$. Consequently the measured value, $n_s \simeq 0.96$, implies $\eta \simeq -0.02$ and so $k \gsim 1$ implies $\epsilon \lsim 0.0008$ and so $r$ is too small: $r \simeq \frac{2}{k^2} \, (n_s - 1)^2 \lsim 0.01$.

For intermediate values of $k$, we see from Figure \ref{rplot} that the value of $r$ depends greatly on the sign of $\eta$. For example, $k=0.1$ gives $r\simeq0.06$ when $\eta<0$, but $r\simeq0.19$ when $\eta>0$. In particular, if error bars were to shrink such that $r=0.2$ is favoured then exponential inflation would predict positive $\eta$, and so a relative sign between $V_0$ and $V_1$.

To be convinced of the reliability of these results two things must be checked. First, it is important to ensure that the slow-roll conditions are satisfied, namely that $\epsilon,|\eta|\ll 1$. Figure \ref{epsetaplot} demonstrates that having $k\simeq0.1$ and $\eta>0$ poses no threat: when $\eta>0$ the slow-roll conditions are met whenever $k \lsim 0.35$. (When $\eta<0$ the conditions are met for any $k$.)

Second, we must ensure that the desired values for $\epsilon$ and $\eta$ are not so large that they imply $e^{-k\phi}$ is too large to drop higher powers of the exponential. In the limit of small $k$ this can be ensured by taking the scale $V_0$ in the potential to be systematically small relative to $V_1$. (Such a condition on $V_0$ is similar in spirit to the choice made for $V_0$ in natural inflation to ensure the potential vanishes at $\phi = 0$.)

\section{Natural large-field inflation}
\label{sec:naturalness}

Although it may be true that a particular classical potential, $V(\phi)$, has a good asymptotic expansion for large $\phi$ in powers of $e^{-k\phi}$ or $1/\phi$, an important part of any successful inflationary model is the understanding of why any desired properties of the classical potential (such as the form of its large-field expansion) should be shared by its quantum corrections. More generally, inflationary models rely on the inflaton mass being smaller than the inflationary Hubble scale, $m^2 \ll H^2$, which is problematic given that quantum corrections generically tend to generate large scalar masses (particularly once the inflationary model is embedded into a UV theory that describes other massive non-inflationary degrees of freedom that couple to the inflaton). An important clue towards unraveling how inflation works is enunciating the mechanism that suppresses these destabilizing quantum corrections. It is these issues that largely motivate complicating inflationary proposals beyond the simplest models and exploring their embeddings into UV completions, such as into string theory \cite{StrInfRev, CQ}.

The usual way of controlling quantum corrections is to show that the assumed functional form is protected by an approximate symmetry \cite{tHooft}, and there are four known kinds of symmetry that can forbid a scalar mass:
\begin{enumerate}
\item Generalized shift symmetries, with the scalar in question a pseudo-Goldstone boson \cite{pGB};
\item Supersymmetry, which ties the scalar mass to fermion masses (that can be protected by chiral symmetries);
\item Scale invariance, which can suppress all dimensionful quantities (though is often anomalous);
\item Extra-dimensional symmetries; for which the 4D scalar in question is actually a Kaluza-Klein mode of the metric or a gauge field in higher dimensions, and so whose mass is protected by higher-dimensional gauge or general-coordinate symmetries.
\end{enumerate}

In this section we show in detail several ways this can be done for the particular case of exponential, trigonometric and some power-law potentials, showing that they can naturally be protected by symmetries if
\begin{itemize}
\item the inflaton is a pseudo-Goldstone boson for a non-compact symmetry group; or
\item the inflaton is a modulus describing the size of a feature (such as the overall volume) of a higher-dimensional geometry.
\end{itemize}
The remainder of this section addresses each of these cases in turn.

\subsection{Trigonometric potentials}

The oldest approach to protecting the inflaton potential is to use option 1 above and make the inflaton a pseudo-Goldstone boson. `Natural inflaton' \cite{NatInf} provides the simplest and earliest example along these lines, and regards the inflaton as a pseudo-Goldstone boson \cite{pGB} for the breaking of a $U(1)$ symmetry leading naturally to a trigonometric potential of the form
\be \label{eq:natpot}
 V(\phi) = \frac{V_0}{2} \left[ 1 - \cos \left( \frac{\phi}{f} \right) \right]  + \cdots = V_0 \, \sin^2 \left( \frac{\phi}{2f} \right) + \cdots \,,
\ee
where an additive constant has been chosen to ensure $V = 0$ at $\phi = 0$. Such a lagrangian arises at low energies if $\phi/f = \vartheta$ arises as the phase of an order parameter, $\Phi(x) = \rho(x) \, e^{i \vartheta(x)}$ with $\langle \rho \rangle \propto f \ne 0$, on which the approximate $U(1)$ symmetry acts as $\Phi \to e^{i\omega} \Phi$.

What is important here is the $U(1)$ symmetry acts on $\phi$ by shifting: $\phi \to \phi + \omega f$, where $\omega$ is the constant symmetry parameter.\footnote{Such a shift symmetry must be spontaneously broken -- {\em i.e} cannot preserve the vacuum -- because the field $\phi$ in particular must take different values in the vacuum before and after a symmetry transformation is performed.} If this had been an exact symmetry of the theory then $\phi$ could only appear in the lagrangian differentiated, $\cL = \cL(\partial_\mu \phi)$, and in particular the scalar potential would have to be $\phi$-independent. But if the symmetry is assumed to be only approximate, explicitly broken by some part of the theory whose energy density, $V_0$, is much smaller than the other scales in the problem, then $V$ can depend on $\phi$ but only by an amount suppressed by the scale $V_0$. The potential takes the form given above if the small terms in the lagrangian that break the symmetry (with size $V_0$) are assumed to transform under the symmetry with charge $\pm 1$: {\em i.e.} $\delta \cL \to e^{\pm i \omega}  \delta \cL$, since this is properly captured in the low-energy $\phi$ theory by terms of the form $V_0 \, e^{\pm i \phi/f}$. The ellipses in eq.~\pref{eq:natpot} involve terms involving higher harmonics --- {\em e.g.} $e^{\pm 2i\phi/f}$ --- which must be proportional to $V_0^2$ and so are suppressed compared with those shown. What is crucial is that because any quantum corrections to $V(\phi)$ must also break the symmetry they must themselves also be proportional to $V_0$, thereby ensuring they are not larger than the form initially assumed for $V$ classically.

\subsubsection*{Nonabelian variants}

Since the abelian theory is perhaps too simple to display the underlying geometrical constructions at work, it is instructive to work through a simple non-abelian examples. Appendix \ref{app:sphere} provides the simplest such example, wherein the broken group is $SO(3)$ and the pseudo-Goldstone bosons parameterize the coset $SO(3)/SO(2)$ --- {\em i.e.} a 2-sphere --- in the symmetry limit. It is often useful to describe instead the complex cover of this space, $SU(2)/U(1)$. The resulting symmetry-breaking potential closely resembles the forms found in chiral perturbation theory for pions \cite{SoftPions, pGB}.

In this case choosing the simplest order parameter for explicit $SO(3)$ breaking gives the following lagrangian
\be
 -  \frac{\cL}{\sqrt{-g}} = F^2 \, \frac{ \partial^\mu z \, \partial_\mu z^*}{(1 + |z|^2)^2} + V_0 - V_1 \left( \frac{1-|z|^2}{1+|z|^2} \right) \,,
\ee
which can be recognized as trigonometric once the complex projective coordinates are re-expressed in terms of the traditional angular coordinates on the sphere,
\be
 z := e^{i\varphi} \, \tan \frac{\theta}{2} \,.
\ee
As is well known, obtaining a successful inflationary slow roll in these models typically requires $F \gsim M_p$, ensuring weaker-than-gravity couplings for the scalar fields.

\subsubsection*{Large-field power laws}

The requirement $F \gsim M_p$ allows the trigonometric potentials (and many others) also to provide useful examples of large-field power law potentials, such as the benchmark form $V \propto \phi^2$ that seems to work so well. The simple power-law form generically obtains when Taylor expanding the potential in a small-field limit, whose validity requires $\phi \ll F$. (For instance for the non-abelian example the quadratic form arises when $|z| \ll 1$ (and, as shown in Appendix \ref{app:sphere}, also when $|z| \gg 1$). But when $F$ is very large this nominally small-field regime is large enough to include what is for inflation the {\em large-field} case: $\phi \gg M_p$. In this way we can see large-field power-law inflation be protected by pseudo-Goldstone symmetries.

\subsection{Exponential potentials}

However pseudo-Goldstone bosons need not only generate trigonometric potentials. After all, it is ultimately the compactness of the underlying $U(1)$ group that identifies $\phi$ and $\phi + 2 \pi f$ and so ensures $V(\phi)$ is periodic in the case of natural inflation. But this compactness is not essential for protecting the inflaton mass from quantum corrections. In this section we show how pseudo-Goldstone bosons for non-compact symmetries often have exponential potentials.

The simplest non-compact symmetry would also be abelian, such as would be obtained if there were a field $\Phi$ for which the symmetry acts as a rescaling, $\Phi \to e^\omega \, \Phi$, rather than the phase rotation considered above. In this case $\Phi$ can be real and nothing requires the corresponding Goldstone field --- for instance, $\log (\Phi/\langle \Phi \rangle)$ --- to be periodic. If we define the symmetry transformation to act on $\phi$ as $\phi \to \phi + \omega f$ as before, and if $\phi$ is the only field that transforms,\footnote{This can normally always be ensured by absorbing an appropriate power of $e^{\phi/f}$ into any other fields that initially do transform.} then we learn again that invariance implies $\cL$ can depend on $\phi$ only through the combination $\partial_\mu \phi$, and so any invariant potential is again independent of $\phi$. If, however, the lagrangian contains small symmetry-breaking terms proportional to $V_1$ transforming as $\delta \cL \to e^{-\omega} \, \cL$, then we expect a small symmetry breaking potential of the form
\be
 V(\phi) = V_0 + V_1 \, e^{-\phi/f} + \cdots \,,
\ee
where, as before, the ellipses include higher orders in $V_1 e^{-\phi/f}$.

\subsubsection*{The $SU(1,1)/U(1)$ model}

Abelian examples are not complicated enough to properly illustrate the difference between the potential for compact and non-compact symmetries, so we next examine the simplest non-abelian non-compact model. This is the two-dimensional hyperbola, $SO(2,1)/SO(2)$, or its complex cover $SU(1,1)/U(1)$, some of whose geometrical properties and definitions are collected in the Appendix \ref{app:hyperb}. As described in the appendix, this space can be described by a complex field, $S$, and the most general generally covariant lagrangian describing the self-interactions of Goldstone bosons for the spontaneous symmetry breaking pattern $SU(1,1) \to U(1)$ (up to two derivatives) has the form
\be \label{eq:Linv}
 \cL_{\rm inv} = - \sqrt{-g} \, \left[ F^2 \, \frac{\partial^\mu S \, \partial_\mu S^*}{(S+S^*)^2} + V_0 \right] \,,
\ee
where $F^2$ and $V_0$ are real constants (both of which we take to be positive). This action has a three-parameter $SU(1,1)$ symmetry of the form
\be \label{eq:SU11trans}
 \delta S = i \alpha + \beta S + i \gamma S^2 \,,
\ee
where $\alpha$, $\beta$ and $\gamma$ are arbitrary constant real parameters.

In order to generate a non-trivial scalar potential we now imagine supplementing eq.~\pref{eq:Linv} with a small symmetry breaking term, proportional to an energy density $V_1$. Following the same steps used above (and in chiral perturbation theory for pions) \cite{pGB} we first specify how the symmetry-breaking perturbation transforms under the group $SU(1,1)$. For illustrative purposes we choose one of the simplest transformation properties: a 3-dimensional representation, corresponding to the 3-vector representation of the 3D Lorentz group $SO(2,1)$.

As shown in Appendix \ref{app:hyperb}, the shape of the resulting potential depends on whether the 3-vector of interest is timelike, spacelike or null. In the case of a timelike order parameter the symmetry-breaking potential turns out to be
\be \label{eq:SpotTL}
 V_{\rm tl} = V_0 - V_1 \left( \frac{1 + 4 \, S^* S}{2(S + S^*)} \right) \,,
\ee
and so is an example of a potential that diverges for some regions of field space (in this case when $S \to \infty$ and $S \to 0$). Choosing a null 3-vector order parameter instead gives a potential that is bounded,
\be \label{eq:Spotnull}
 V_{\rm null} = V_0 - \frac{V_1}{S+S^*} \,.
\ee

Of course, breaking the symmetry modifies the scalar kinetic terms in addition to generating a non-trivial scalar potential. However unlike for the potential the invariant lagrangian, $\cL_{\rm inv}$, already had a non-trivial invariant metric and so it is the invariant part of this metric that dominates the small symmetry-breaking contributions. The situation is different (during inflation) for the scalar potential because the `force', $\partial V/\partial S$, strictly vanishes in the symmetry limit, and so is dominated by any small symmetry-breaking effects. We are led in this way to a natural pseudo-Goldstone inflaton lagrangian of the form
\be
 \cL = - \sqrt{-g} \, \left[ F^2 \, \frac{\partial^\mu S \, \partial_\mu S^*}{(S+S^*)^2} + V \right] \,,
\ee
with $V$ given in eq.~\pref{eq:SpotTL} or \pref{eq:Spotnull}.

In terms of the real and imaginary parts, $S := s + i a$, the lagrangian obtained using the potential eq.~\pref{eq:Spotnull} becomes
\be
 \cL = - \sqrt{-g} \, \left\{ F^2 \left( \frac{\partial^\mu s \, \partial_\mu s + \partial^\mu a \, \partial_\mu a}{4s^2} \right) + V_0 -  \frac{V_1}{2s}  \right\} \,,
\ee
whose potential (due to the axionic shift symmetry of $a$) always has a stationary point with constant $a$. Because of this symmetry it is always possible to find an inflationary solution for which only $s$ evolves. (Of course $a$ can contribute non-trivially to primordial fluctuations to the extent that it is generically lighter than the inflationary Hubble scale, a point to which we return below.) Inspection of the kinetic term shows that the canonical variable is
\be
 \phi = \frac{F}{\sqrt 2} \; \ln \, s \,,
\ee
and so for large $s$ the effective inflationary potential has an exponential form,
\be
 V(\phi) \simeq V_0 - \frac{V_1}{s} \simeq  V_0 - V_1 \, e^{-k \phi/M_p} \qquad \hbox{with} \qquad
 k = \frac{\sqrt2 \, M_p}{F} \,.
\ee
In this model obtaining $k$ in the small-$k$ regime --- {\em i.e.} $k \lsim 0.3$ --- requires choosing $F \gsim 5 M_p$. Whether this is possible in practice is a question whose answer requires knowing the UV completion that is relevant at energies above the inflationary scale (more about which below).

Besides protecting the inflaton mass, the nonabelian examples show that other scalars can also be kept light during inflation, with possibly observable effects due to the large fluctuations that get imprinted onto them during horizon exit. Such `isocurvature' fluctuations back at the scale of horizon exit need not be dangerous in themselves, but become poison if subsequent evolution feeds them observably into the observed temperature fluctuations of the CMB (for which iso-curvature fluctuations are are strongly disfavoured by observations).

In general this can be a problem for light axion-like fields \cite{QCDAxion} that are present during inflation, even for $F$ somewhat larger than the Planck scale. However, whether they can actually kill a model is model-dependent, since there are also a number of ways to evade these bounds, such as:
\begin{itemize}
\item Make the axion be precisely massless, so as not to have an appreciable energy density in the later universe;
\item Make the axion decay between inflation and the later universe, or otherwise equilibrate its energy density into the thermal bath that later dominates the later universe;
\end{itemize}
and so on.

\subsection{Beyond pseudo-Goldstone inflatons}

There is another natural class of theories that lead to exponential inflation, for which the form of the potential is protected by higher-dimensional symmetries rather than 4D generalized shift symmetries. For the simplest examples the inflaton arises within a higher-dimensional context as a {\em modulus} (such as the volume) describing the geometry of extra dimensions. Because moduli parameterize the classical vacua of extra-dimensional models, at the leading approximation they are massless: corresponding to massless Kaluza-Klein (KK) modes of the extra-dimensional metric.

In general moduli acquire masses, either through classical interactions or quantum effects. Often these masses are small compared to generic KK scales, making them natural candidates for a light inflaton. In such cases the symmetry that protects the form of the 4D scalar potential can be higher-dimensional general covariance (and/or supersymmetry), since (for large dimensions) this forces the energy of the system to be organized into an expansion in powers of curvatures and their derivatives, which often becomes a series in inverse powers of the modulus of interest \cite{uber}. Such symmetries are not easily described simply in 4D terms as generalized shift symmetries for pseudo-Goldstone bosons.

To be concrete, consider the following simple example \cite{rasinflaton}. Suppose the extra-dimensional metric has the form
\be
 \exd s^2 = \tilde g_{\ssM\ssN} \exd x^\ssM x^\ssN = \hat g_{\mu\nu}(x) \, \exd x^\mu \exd x^\nu + \rho^2(x) \, g_{mn}(y) \, \exd y^m \exd y^n \,,
\ee
in a theory involving $4+n$ spacetime dimensions. In this case the higher-dimensional Ricci curvature scalar is given by
\be
 \tilde R_{(4+n)} := \tilde g^{\ssM \ssN} \tilde R_{\ssM \ssN} = \hat R_{(4)} + \frac{R_{(n)}}{\rho^2} + \frac{2n}{\rho} \, \hat \Box \rho + \frac{n(n-1)}{\rho^2} \, \hat g^{\mu\nu} \partial_\mu \rho \partial_\nu \rho \,,
\ee
and so (after integrating by parts) the higher-dimension Einstein action dimensionally reduces to give
\bea
 \cL_{\rm EH} &=& - \frac{M_p^2}{2} \sqrt{-\hat g} \left( \frac{\rho}{L} \right)^n \hat g^{\mu\nu} \left[ \hat R_{\mu\nu} - \frac{n(n-1)}{\rho^2} \, \partial_\mu \rho \, \partial_\nu \rho \right] + \cdots \nn\\
 &=&  - \frac{M_p^2}{2} \sqrt{- g} \; g^{\mu\nu} \left[ R_{\mu\nu} + \frac{n(n+2)}{2\rho^2}  \, \partial_\mu \rho \, \partial_\nu \rho \right] + \cdots \,,
\eea
where $L$ denotes the vacuum value of the extra-dimensional radius and the last line transforms to Einstein frame, $\hat g_{\mu\nu} = (L/\rho)^n g_{\mu\nu}$.

In particular the kinetic term for $\rho$ shows that $\rho$ is related to the canonically normalized variable, $\phi$, by
\be
 \rho = \ell \, e^{\lambda \phi/M_p} \qquad \hbox{where} \qquad
 \lambda^2 = \frac{2}{n(n+2)} \,.
\ee
Notice that the values of $\lambda$ for the main choices for $n$ (listed in Table 1) can give comparatively small $\lambda$.

\vspace{2mm}
\begin{center}
\begin{tabular}{c||c|c|c|c|c|c|c}
  $n$ & 1 & 2 & 3 & 4 & 5 & 6 & 7 \\
  \hline
  $\lambda$ & 0.82 & 0.50 & 0.37 & 0.29 & 0.24 & 0.20 & 0.18 \\
\end{tabular}
\vspace{4mm}\\
Table 1: $\lambda$ as a function of number of extra dimensions.
\end{center}
\vspace{2mm}

The potential energy for $\rho$ is model-dependent, but typically general covariance requires it to arise as a curvature expansion, and so as a series in $1/\rho^2$ (at least for large $\rho$, which is the domain of validity of semiclassical calculations), {\em e.g.}
\be
 \cL_{\rm pot} =  \sqrt{-\hat g} \left( \frac{\rho}{L} \right)^n \Bigl[ A + B \, \tilde R_{(4+n)} + \cdots \Bigr]
 = \sqrt{- g} \; \left( \frac{L}{\rho} \right)^n \left[ A + \frac{ B \, R_{(n)}}{\rho^2} + \cdots \right] \,,
\ee
clearly generating an exponential potential once expressed in terms of the canonical variable. In the particular case where it is a higher-dimensional cosmological constant that dominates its $\rho$ dependence we would find $k = \lambda n = \sqrt{2n^2/n(n+2)}$, which approaches $\sqrt2$ for $n$ large. This will not provide the small values of $k$ required to obtain  $r\sim 0.1$.

There are two challenges that these models usually face. One is obtaining an approximately constant term, $V_0$, in the potential. This is often difficult because of the propensity for these models to admit extra-dimensional flat space as a solution. It is not impossible to arrange, however, typically by having multiple moduli, of which some are hung up in a meta-stable vacuum \cite{KMInf, Fibre, EvadS}.

The second general challenge in making extra-dimensional inflationary models is having control over the potential for stabilizing {\em all} of the moduli. A convincing case for inflation requires a mechanism for stabilizing the extra dimensions because to show that inflation happens one must argue that the shallow inflationary direction is the {\em steepest} direction available for the system's evolution. So it is insufficient just to find a single degree of freedom that gives a slow roll when considered in isolation; one must as well show that there are no other directions in field space that are steeper. This turns out to be much more difficult and is best explored within a string inflationary context, where it has only been possible for the last decade or so \cite{KKLMMT, StrInfRev}\, only after concrete mechanisms of moduli stabilisation were developed \cite{GKP,KKLT,LVS}. (See also \cite{6DInf} for a non-string example, including modulus stabilization.)

\subsection{UV completions}

As is often the case inflationary models such as those considered above raise questions whose addressing requires embedding into some sort of UV completion, and because $r \gsim 0.1$ implies the inflationary scale is now known to be quite high this UV completion usually involves some formulation of quantum gravity. (In what follows we take this to be string theory, since this is the best developed formulation in which questions can be asked in their crispest form.) We briefly consider some of the questions here.

\subsubsection*{Global symmetries}

One question starts with the observation that models like natural inflation and natural exponential inflation treat the inflaton as a pseudo-Goldstone boson for an approximate global symmetry. But it is widely believed that global symmetries cannot exist in quantum gravity \cite{NoGlobSinGrav}, and this is known to be true in particular for string theory \cite{NoGlobSinST}. How can global symmetries be there to play a role protecting the inflaton in the lower-energy world?

It turns out on closer examination that although exact global symmetries are impossible, nothing in string theory precludes the existence of {\em approximate} global symmetries \cite{StrFlav}. In specific examples this happens together with the existence of weak-than gravity couplings (such as happens for Goldstone bosons when $F > M_p$), similar to recent conjectures made in another context \cite{WeakerThanGrav}.

\subsubsection*{Large decay constants}

We see from the above that both conventional `trigonometric' natural inflation and exponential natural inflation (as considered here) need a decay constant larger than $M_p$, although this requirement arises differently in the two cases. For trigonometric natural inflation the requirement $f \gg M_p$ arises as a consequence of the slow-roll conditions themselves, since otherwise $\epsilon$ and $\eta$ are generically not sufficiently small. In exponential inflation, by contrast, slow roll automatically occurs for sufficiently large $\phi$, and it is instead the phenomenological requirement $r \gsim 0.1$ that demands $F \gsim M_p$.

Having trans-Planckian values for $f$ or $F$ implies $\phi$ couplings that are weaker than gravitational strength and one might ask whether this raises naturalness problems in itself, given that everybody couples to gravity. As has been argued elsewhere, this need not be a problem to the extent that the fundamental scale associated with gravity is below the 4D Planck scale \cite{uber}, such as happens in extra-dimensional models for whom the extra dimensions are relatively large.\footnote{The BICEP2 results tell us that the gravity scale cannot be {\em too} low at the epoch of horizon exit, but still allow a low gravity scale at later epochs if the extra-dimensions themselves grow in size during or after inflation \cite{XDgrow}.}

But although there need not be a problem of principle in having $F \gsim M_p$, a crisp enunciation of the issues usually requires having a UV completion for the inflationary model of interest. Among the attractive features of inflationary models is that they arise within a variety of sensible UV completions (such as string theory) \cite{StrInfRev, CQ}, allowing these questions to be cast in a usefully precise form. Experience with these shows that obtaining large decay constants is difficult, though it is also true that no no-go theorems exist. We next briefly discuss two related types of known UV completions.

\subsubsection*{4D supergravity and Moduli Spaces}

4D supergravity provides one class of examples for which the $SU(1,1)/U(1)$ coset bosons, and their generalizations to other groups, have long been known robustly to arise \cite{Sugrasigma, LindeSU11}.

In these supergravities the $SU(1,1)/U(1)$ invariant kinetic terms for the complex field $S$ often arise through an additive contribution to the K\"ahler function of the form
\be
 K(S,S^*, \chi^a, \chi_a^*) = \cK(\chi^a, \chi_a^*) - \ln( S + S^*) \,,
\ee
where $S$ is the complex scalar for a chiral multiplet. This kind of contribution has long been known to arise in particular among the 4D supergravities that capture the low-energy limit of heterotic \cite{hetero} and type IIB \cite{TypeIIB} string vacua.

Inflationary models of this type have recently been re-examined \cite{InflationSugra} given the great success\footnote{These models do less well in view of the BICEP2 data, because of their focus on the regime $k \gsim \cO(1)$. It is not clear if UV complete models can be constructed with $k\ll 1$ that give rise to large tensor modes as discussed before.} that exponential potentials  had in describing the primordial fluctuations seen by the Planck collaboration \cite{Planck}. The success of models of this type is very natural given that similar models are also known for UV completions in string theory for which this type of supergravity provides the low-energy 4D limit. Indeed, historically it was the UV completions that were found first for the exponential \cite{Fibre} and related \cite{KMInf} models.

We next compute examples of scalar potentials that are consistent with the pseudo-Goldstone symmetries considered above. We do so with two goals in mind: to show some of the ways that supersymmetry can supplement the protection of pseudo-Goldstone boson potentials; and to describe a completely different class of symmetry breaking than considered above that also generates exponential potentials. To that end we compute below the $D$-term and $F$-term potentials that naturally arise in supersymmetric versions of the $SU(1,1)/U(1)$ model. (Details may be found in Appendix \ref{app:susy}.)

We start by considering the symmetries of the following K\"ahler potential,
\be
 K = -p \log(S+S^*) \,.
\ee
which is invariant (up to a K\"ahler transformation) under the three isometries of eq.~\pref{eq:SU11trans}, under which
\be
 \delta K = -p \beta  -i p \gamma (S- S^*) \,.
\ee
Comparing to $\delta K = r(S) + \hbox{c.c.}$ we can read off
\be \label{rAs}
 r(S)  = - \frac{p \beta}{2}  -ip\gamma \,S \,,
\ee
which is unique up to the addition of an imaginary constant $-i \xi$.

\medskip\noindent{\it $D$-term Potential}

\medskip\noindent
For simplicity assume the first of these symmetries is gauged in a supersymmetric way using the gauge fields $A_\mu^\ssA$. Furthermore, assume imaginary shifts of $S$ are anomalous so it is the real part of $S$ that plays the role of the gauge coupling (as is often the case), so the gauge kinetic term is (see Appendix \ref{app:susy} for the result when the other generators are gauged)
\be
 \cL_{k\ssA} = -\frac14 \, {\rm Re} f(S) \, F_{\mu\nu}^\ssA F^{\mu\nu}_\ssA \,,
\ee
with $f(S) = S$.  The corresponding $D$-term potential becomes
\be
 V_\ssD = \frac{\cP \cP^*}{2 \, {\rm Re} f(S)}
\ee
where the moment map $\cP$ is given by
\be
 \cP = i \big(k^\ssS \pd_\ssS K - r \big) = - \pd_\ssS K = \frac{p}{S+S^*} \,.
\ee
Substituting, we find the $D$-term potential
\be
 V_\ssD(s) = \frac {p^2 M_p^4}{8s^3} \,.
\ee

The axion in this theory turns out to be eaten by the gauge field, after which we find the inflaton lagrangian in this case gives
\be
 - \frac{\cL_s}{\sqrt{-g}} = \frac{ p M_p^2}{4s^2} \; \pd_\mu s \pd^\mu s + V_\ssD(s) \,.
\ee
This again gives an exponential potential, $V = V_1 \, e^{-k\phi/M_p}$ once written in terms of the canonical field,
\be
 \phi:=\sqrt{\frac 2p} \, \ln s \,,
\ee
and so $V_1 = p^2 M_p^4/8$ and $k = 3\sqrt{p/2}$. Requiring $k = 0.1$ gives $p = 1/450 \simeq 0.002$.

\medskip\noindent {\it $F$-term Potential}

\medskip\noindent

Invariance of the superpotential under imaginary shifts of $S$ allows
\be
 W = W_0 \, e^{a S} \,,
\ee
where $a = 0$ if this shift is not an $R$ symmetry (or if we work to finite order in the gauge coupling, $1/s$). In this case the $F$-term potential must take the form
\be
V_\ssF = |W_0|^2 e^K \big(g^{SS^*} \pd_S K \pd_{S^*} K -3 \big) = (p-3) \frac{|W_0|^2}{(S+ S^*)^p} \,,
\ee
and so is again exponential once written in terms of the canonical variable.

\subsection*{String compactifications}

Coset-space constructions arise quite generally in the moduli spaces that arise within string-theoretic models. The model-independent dilaton fits precisely into the specific $SU(1,1)/U(1)$ coset described above and complex-structure and K\"ahler moduli of particular compactifications are often given by larger coset spaces $G/H$ or their subsets \cite{louis}. The continuous approximate symmetries of these cosets are not inconsistent with the absence of global symmetries however because they are broken to the standard duality symmetries by quantum effects, thereby lifting the flatness of the potential for the corresponding moduli spaces after supersymmetry breaking. In some cases the continuous symmetry may remain as an approximate global symmetry of the effective field theory \cite{StrFlav}.

But UV completions are also useful in that they permit one to move past `generic' statements about low-energies by allowing explicit calculations of the low-energy action. What is interesting is that sometimes this action turns out not to be completely generic, and give rise to light moduli in unusual ways. In particular, accidental global symmetries can arise accidentally in an unexpected fashion.

For example, light moduli can robustly arise in string compactifications with more than one K\"ahler modulus. This can be seen most clearly for type IIB vacua for which the tree-level K\"ahler potential depends just on the Calabi-Yau volume $\vo$:\footnote{The tree-level $K$ also depends only on $\vo$ for heterotic vacua.}
\be
 K_{\rm tree} = -2\ln\vo\,,
\label{Ktree}
\ee
where $\vo$ is here regarded as a function of the real parts of the complex fields that label the various moduli in supersymmetric vacua. This implies in particular that $K_{\rm tree}$ does not depend on the axions corresponding to the imaginary parts of these fields. Furthermore, the fact that the moduli have no potential at all at leading order is expressed in the low-energy 4D effective theory by having the leading-order superpotential not depend on the moduli, and by having $\vo$ arise as a homogeneous function of specific degree. These two conditions imply that the resulting potential has the `no-scale' form \cite{noscale}, and so precisely vanishes. Supersymmetric non-renormalization theorems also imply the moduli do not arise in the corrections to the superpotential to any order in perturbation theory (though they can appear in new ways in the perturbative corrections to $K$).

Having $K_{\rm tree}$ depend only on $\vo$ turns out to imply a hierarchy of masses among the low-energy moduli, with the lighter moduli being natural inflaton candidates \cite{Fibre, CQ}. Part of this hierarchy (such as the suppression of axion masses) is not surprising in itself, since it is due to the presence of the axionic shift symmetries (which are what precludes them from appearing in $K_{\rm tree}$). However, the pattern of light moduli masses is richer than just having light axions. If the total number of moduli is $n$, then the fact that $K_{\rm tree}$ depends only on one of them, $\vo$, means that the other $(n-1)$ moduli orthogonal to $\vo$ only receive masses once $\alpha'$ or string-loop corrections to $K_{\rm tree}$ are included \cite{Berg}. All of these moduli enjoy an effective accidental shift symmetry provided one works at leading order, where $K \simeq K_{\rm tree}$, with symmetry-breaking effects generated by subleading $\alpha'$ and quantum corrections (which generically mix all the moduli).

These effective approximate shift symmetries protect the inflaton
potential against higher dimensional operators, but they are insufficient
to guarantee that all symmetry-breaking corrections are small enough not
to ruin an inflationary picture. There are dangerous kinds of corrections that in principle can arise, but explicit calculations show do not. Their absence turns out to be ensured by a very interesting interplay of leading-order properties and string loop effects \cite{CQ}:
\begin{enumerate}
\item The tree-level scalar modulus potential derived from (\ref{Ktree}) is of the \emph{no-scale} type (as mentioned above), and so to leading order is precisely flat for all the K\"ahler moduli;
\item The loop effects enjoy an extended no-scale structure \cite{ExtNoScale}, that also suppresses the leading order loop contribution to the scalar potential;
\item $\alpha'$ corrections to $K$ are therefore the leading effect that breaks the no-scale structure \cite{alpha}, but lift only the volume direction, $\vo$;
\item All of the $(n-1)$ directions orthogonal to $\vo$ are thus flatter and behave as good inflaton candidates (these directions can be lifted by subleading loop \cite{Fibre} or non-perturbative effects \cite{KMInf,Poly}).
\end{enumerate}
At present it is not known how to capture all of these features purely in terms of symmetry properties of the low-energy theory, though there is much interest in doing so given that they appear to be stable against corrections and so are technically natural.

\section{Conclusions}
\label{sec:conclusions}

This seems to be the decade of fundamental scalar fields in particle physics. If confirmed, the results of BICEP2 open a golden window onto the high energies relevant to UV completions of the standard models of particle physics and cosmology. The fact that these results point to high energies, of order the GUT scale, hints at new physics close to the Planck scale. Having the simplest slow-roll inflation successfully capture CMB observations while also generating sufficiently large tensor modes puts strong constraints on specific inflationary models.

In particular, the indication that trans-Planckian values of the scalar fields may have been explored is likely to be quite informative. Although these need not be problematic for a controlled EFT approach (provided the large-field energy densities are not too large), there is little in our experience at small fields that is guaranteed to go over as well to the properties of physics at trans-Planckian field values.

We argue here a point of view (that we believe is widely appreciated) that a good way to formulate large-field physics in a precise and controlled way is if the theory is close to a symmetric limit that controls the large-field asymptotic forms. Such a symmetry can also protect against large quantum corrections and thereby help control the validity of the EFT of interest.

Turning this argument around, the BICEP2 results may be hinting at the existence of approximate symmetries of whatever UV complete theory applies at the high energies at which we now know inflation must take place. This makes it important to explore systematically the possible symmetries that can play a role in this regard, and in this article we explore several kinds that can be relevant: spontaneously broken approximate global symmetries (for which the inflaton is a pseudo-Goldstone boson), supersymmetry and extra-dimensional spacetime symmetries.

For pseudo-Goldstone inflatons, we are led to new kinds of inflaton potentials in addition to the standard shift/axionic symmetries that have long been studied. By studying both non-abelian and non-compact coset spaces we identify several kinds of natural potentials, including both a class of exponential potentials. These exponential potentials can describe observations well, including allowing $r \gsim 0.1$. They make it difficult to obtain $r$ too much larger (such as being as large as 0.2) when $n_s \simeq 0.96$, and so could easily become ruled out as the errors on $r$ and $n_s$ improve.

We also show how the simplest quadratic potential of chaotic inflation can be included in the class of natural pseudo-inflaton models. This becomes possible because pseudo-Goldstone inflation allows the potential to be Taylor expanded for small fields whenever $\phi \ll F$ (generically leading to quadratic behaviour near a minimum), but this small-field regime can include trans-Planckian fields when the decay constant satisfies $F > M_p$ (as it typically does in inflationary models).

Finally, we indicate how these natural potentials (including in particular exponential inflation) often emerge from UV completions, and in particular are generic in supergravity and string theory. This helps to have a broader perspective in the search for concrete UV complete models of inflation. Finding a fully UV satisfactory realization of all features of inflation, including modulus stabilization remains an open question.

\section*{Acknowledgements}

We thank Mina Arvanitaki, Joe Conlon, Ross Diener, Doddy Marsh, Leo van Nierop, Roberto Valandro, Giovanni Villadoro, Wei Xue and Itay Yavin for useful discussions. Our research was supported in part by funds from the Natural Sciences and Engineering Research Council (NSERC) of Canada. Research at the Perimeter Institute is supported in part by the Government of Canada through Industry Canada, and by the Province of Ontario through the Ministry of Research and Information (MRI).

\appendix

\section{Formulations invariant under field redefinitions}
\label{app:fieldredef}

In this appendix we briefly review the formulation of the slow-roll criteria in a way that is invariant under field redefinitions. Such a formulation is useful inasmuch as it allows the rephrasing of what large-field inflation means in a way that is not tied to a specific set of target-space coordinates.

In general the inflaton sector involves many fields, $\phi^a$, whose most general interactions in a derivative expansion have the form
\be
 - \frac{\cL}{\sqrt{-g}} = V(\phi) + \frac12 \, \cG_{ab}(\phi) \, \partial^\mu \phi^a \, \partial_\mu \phi^b \,,
\ee
where the scalar potential, $V$, transforms under field redefinitions, $\delta \phi^a = \xi^a(\phi)$ as a scalar field, $\delta V = \xi^a \partial_a V$, while the target-space metric, $\cG_{ab}$, transforms as a symmetric (and positive-definite) rank-2 covariant tensor: $\delta \cG_{ab} = \xi^c \partial_c \cG_{ab} + \cG_{ac} \partial_b \xi^c + \cG_{cb} \partial_a \xi^c$.

The usual formulation of slow-roll parameters in terms of derivatives of $V$ assume a flat target-space metric, $\cG_{ab} = \delta_{ab}$. Although this can always be arranged in the immediate vicinity of any specific field, $\phi_\star^a$, by going to Gaussian normal coordinates at this point, it cannot be arranged everywhere throughout a region in the target space unless this region is flat; {\em i.e.} the Riemann tensor built from $\cG_{ab}$ vanishes there: ${\cR^a}_{bcd} = 0$.

In general inflation can be described using normal coordinates provided the slow roll itself doesn't carry $\phi^a$ too far from its starting point, $\phi_\star^a$. This requires the target-space proper distance between the initial and final field configurations,
\be
 \sigma(\phi_\star, \phi) = \int_C \exd s \; \sqrt{ \cG_{ab} \, \frac{\exd \phi^a}{\exd s} \, \frac{\exd \phi^b}{\exd s}} \,,
\ee
not to be large compared with the local radii of target-space curvature defined by ${\cR^a}_{bcd}$. Here the integration is along the curve, $C$, defined by integrating the scalar field equations,
\bea
 0 &=& \cG_{ab} \, g^{\mu\nu} \, D_\mu \partial_\nu \phi^b  - \partial_a V \nn\\
 &=&  \cG_{ab} \, g^{\mu\nu} \left[ \partial_\mu \partial_\nu \phi^b + \cC^b_{cd} \, \partial_\mu \phi^c \, \partial_\nu \phi^d - \Gamma^\lambda_{\mu\nu} \, \partial_\lambda \phi^b \right] - \partial_a V  \,,
\eea
where $\cC^a_{bc}(\phi)$ is the Christoffel symbol built from $\cG_{ab}(\phi)$ and its first derivatives, while $\Gamma^\lambda_{\mu\nu}$ is the Christoffel symbol built from the 4D spacetime metric, $g_{\mu\nu}$. Restricted to a homogeneous roll in time this becomes
\be
 \cG_{ab}(\phi) \left( \frac{\exd^2 \phi^b}{\exd s^2} + \cC^b_{cd}(\phi) \, \frac{\exd \phi^c}{\exd s} \, \frac{\exd \phi^d}{\exd s}  + 3 H \, \frac{\exd \phi^b}{\exd s} \right) - \partial_a V(\phi) = 0 \,,
\ee
which becomes a target-space geodesic when $\partial_a V$ vanishes. Here $H$ is the Hubble scale of the 4D spacetime metric, $g_{\mu\nu}$.

In terms of these quantities the slow-roll approximation occurs whenever the quantity $\ddot \phi^a + \cC^a_{bc} \, \dot \phi^b \, \dot \phi^c$ is negligible compared with $3 H \dot \phi^a$. In this case the field equations imply $\dot \phi^a$ points in the direction set by the gradients of the potential, $\cG^{ab} \partial_b V$.

Necessary conditions for slow roll to be a good approximation are when $\epsilon$ and $\eta$ are small, where
\be
 \epsilon := \frac12 \, \cG^{ab} \, \partial_a V \, \partial_b V \,,
\ee
generalizes $\epsilon$ as defined in eq.~\pref{eq:epsetadef}. When $\epsilon = 0$ then $\eta$ generalizes to the most negative of the eigenvalues, $\lambda$, defined by the eigenvalue equation
\be
 \cG^{ab} \Bigl( \partial_b \partial_c V - \cC^d_{bc} \, \partial_d V \Bigr) v^c = \lambda v^a \,,
\ee
since this generalizes the most negative eigenvalue of the second-derivative matrix of $V$, but does so in a way that ensures $\lambda$ transforms as a scalar under field redefinitions. When $\epsilon \ne 0$ what counts is the second derivative of the potential in the direction of the slow roll, so $\eta$ instead generalizes to
\be
 \eta := \cG^{ad} \cG^{be} \Bigl( \partial_a \partial_b V - \cC^c_{ab} \, \partial_c V \Bigr) \frac{\partial_d V \partial_e V}{2\,\epsilon} \,.
\ee

Finally, the redefinition-invariant version of the Lyth bound is the statement that the invariant distance, $\sigma(\phi_\star, \phi)$, becomes trans-Planckian once measured between horizon exit and inflation's end.

The utility of these definitions is that they also apply after a field redefinition is used to transform asymptotically large fields to some finite value. For instance, for a complex field $z$ the transformation $w = 1/z$ takes infinite values of the field $z$ to vanishing values for $w$. If inflation occurs at large field, $z \to \infty$, then it is mapped to small fields, $w = 0$, allowing large-field models to be translated into small-field models (but with target-space metrics that ensure that the same large-field physics nonetheless arises). Having $V$ diverge like a power of $|z|$ for large $z$ means that $V$ must also diverges as $w \to 0$ like the same power of $1/|w|$. Similarly, if $V$ is bounded for large $z$ it also remains bounded for $w \to 0$. On the other hand, if the large-$z$ metric is Cartesian, $\exd \sigma^2 = \exd z \, \exd z^*$, then the metric near $w = 0$ is $\exd \sigma^2 = \exd w \, \exd w^*/|w|^4$, making the invariant distance $|w^{-1} - w_\star^{-1}|^2$ rather than simply $|w - w_\star|^2$.

\section{Some simple coset spaces}
\label{app:coset}

In this appendix we collect some formulae expressing properties of the simplest non-abelian coset spaces: the compact space $SU(2)/U(1)$ (the doubly covered two-sphere) and the noncompact space $SU(1,1)/U(1)$ (the doubly covered hyperbolic plane).

\subsection{The sphere $SO(3)/SO(2)$}
\label{app:sphere}

We start with the 2-sphere, $SO(3)/SO(2)$, which can be regarded as the region of fixed proper radius in three-dimensional Eucliean space:
\be \label{eq:sphdef}
 u^2 + x^2 + y^2 = \ell^2  \,.
\ee
The standard spherical coordinates for this 2D surface is given by $\{\theta, \varphi \}$ with $0 \le \theta < \pi$ and $0 \le \varphi < 2\pi$, where
\be \label{eq:sphcoorddef}
  u= \ell \cos \theta \,, \qquad x = \ell \sin \theta \cos \varphi \quad \hbox{and} \quad y = \ell \sin \theta \sin \varphi \,,
\ee
since these satisfy eq.~\pref{eq:sphdef} identically. Complex projective coordinates for this space are related to these by
\be
 z := \frac{x+iy}{\ell+t} = e^{i\varphi} \, \tan \frac{\theta}{2} \,,
\ee
where $0 \le |z| < \infty$.

In terms of these coordinates the induced metric inherited from Euclidean space is
\be
 \exd \sigma^2 = \exd u^2 + \exd x^2 + \exd y^2 = \ell^2 \Bigl( \exd \theta^2 + \sin^2 \theta \, \exd \varphi^2 \Bigr) = 4\ell^2 \, \frac{ \exd z \, \exd z^*}{(1 + |z|^2)^2} \,.
\ee

\subsubsection*{`Large-field' limits}

We have the following pseudo-Goldstone lagrangian,
\be
 -  \frac{\cL}{\sqrt{-g}} = F^2 \, \frac{ \partial^\mu z \, \partial_\mu z^*}{(1 + |z|^2)^2} + V_0 - V_1 \left( \frac{1-|z|^2}{1+|z|^2} \right) \,,
\ee
for which we can explore various large-field limits. Our goal is to show how `large-field' limits can be mapped to what are geometrically small-field regimes, ultimately because the compactness of the target space ensures that all field configurations are really a small proper distance from all others. Yet the result can still be large-field in the inflationary sense of the Lyth bound, since small fields in the geometrical sense can nonetheless be large --- {\em i.e.} trans-Planckian --- in the inflationary sense. In this particular example the large-field model we obtain has a simple quadratic potential, $V \propto +V_1 |w|^2$, in a controlled approximation at large fields.

To this end consider the limit of large $|z|$. This is most easily explored by mapping infinity to zero using the transformation $z := 1/w$, in terms of which the action becomes
\be
 -  \frac{\cL}{\sqrt{-g}} = F^2 \, \frac{ \partial^\mu w \, \partial_\mu w^*}{(1 + |w|^2)^2} + V_0 + V_1 \left( \frac{1-|w|^2}{1+|w|^2} \right) \,,
\ee
and we are now to explore near $w = 0$. This has the same form as the original lagrangian with the sign of $V_1$ changed. Expanding to leading order gives
\be
 -  \frac{\cL}{\sqrt{-g}} \simeq F^2 \, \partial^\mu w \, \partial_\mu w^* + (V_0 + V_1) - 2 V_1 |w|^2 + \cdots \,.
\ee
In terms of the canonical field, $\zeta = F w$, the domain of validity, $|w| \ll 1$, of this expansion corresponds to $|\zeta| \ll F$, which can include large, trans-Planckian, fields if $F \gg M_p$. Notice also that inflation wants a positive coefficient of the quadratic term, which is what tells us whether inflation occurs near $z = 0$ or $w = 0$.

\subsection{The hyperbola $SO(2,1)/SO(2)$}
\label{app:hyperb}

Consider now a simple noncompact surface, $SO(2,1)/SO(2)$, which can be regarded as one sheet of a spacelike hyperbola defined by fixed timelike proper interval in three-dimensional Minkowski space:
\be \label{eq:hypdef}
 t^2 - x^2 - y^2 = \ell^2 \qquad \hbox{and} \qquad t > 0 \,.
\ee
A convenient set of coordinates for this 2D surface is given by $\{\tau, \theta\}$ where $0 \le \tau < \infty$ and $0 \le \theta < 2 \pi$ with
\be \label{eq:hypcoorddef}
 t = \ell \cosh \tau \,, \qquad x = \ell \sinh \tau \cos \theta \quad \hbox{and} \quad y = \ell \sinh \tau \sin \theta \,,
\ee
since these satisfy eq.~\pref{eq:hypdef} identically.

Complex projective coordinates for this space are related to these by
\be
 z := \frac{x + i y}{\ell + t} = e^{i\theta} \left( \frac{ \sinh \tau}{1 + \cosh \tau} \right) = e^{i\theta} \, \tanh \left( \frac{\tau}{2} \right) \,,
\ee
and so $0 \le |z| < 1$ for $0 \le \tau < \infty$. In terms of these coordinates the induced metric inherited from Minkowski space is
\be
 \exd \sigma^2 = - \exd t^2 + \exd x^2 + \exd y^2 = \ell^2 \Bigl( \exd \tau^2 + \sinh^2 \tau \, \exd \theta^2 \Bigr) = 4\ell^2  \frac{ \exd z \, \exd z^*}{(1 - |z|^2)^2} \,.
\ee

In supersymmetric theories this coset space is often encountered through the alternative complex coordinates,
\be \label{app:Svsz}
 S = \frac12 \left( \frac{1 + z}{1 - z} \right) \qquad \hbox{and so} \qquad
 z = \frac{2S-1}{2S+1} \,,
\ee
for which
$0 \le |z| < 1$ corresponds to $0 \le |S| < \infty$, with $S \to \infty$ corresponding to $z \to 1$. Notice also that
\be
 S + S^* = \frac{1 - |z|^2}{|1 - z|^2}  \qquad \hbox{and} \qquad
 \exd S = \frac{ \exd z}{ (1 - z)^2} \,,
\ee
which allow the target-space metric to be written
\be
 \exd \sigma^2 = 4 \ell^2 \, \frac{ \exd S \, \exd S^*}{(S+S^*)^2} \,.
\ee

\subsubsection*{Scalar potential}

We are interested in constructing potentials for $z$ or $S$ given a simple symmetry-breaking sector. What is new in the noncompact case is the possibility of having points in field space separated by an infinite proper distance, allowing (but not requiring) the simplest potentials to diverge for some parts of field space. We show examples of both types below, but before doing so two generic remarks are in order:
\begin{itemize}
\item Because any dependence of $V$ on the fields is a symmetry-breaking effect, a divergence of $V$ for some fields indicates a strong deviation from the limit where the symmetry is approximate. In such cases care must be taken to maintain control over the approximations in the large-$V$ limit.
\item Any bounded $V$ may be expanded about $z = 1$, and the smoothness of $V$ automatically ensures the existence of an expansion in powers of $1/S$, ultimately leading (as we shall see) to an exponential potential.
\end{itemize}

A simple choice for the symmetry-breaking lagrangian parallels the compact case by assuming the symmetry-breaking parameters transform like a 3-vector, $W_\mu$, of the 3D Lorentz group, $SO(2,1)$. There are then several cases to consider, depending on whether $W_\mu$ is timelike, spacelike or null. Consider for instance the two following examples.

\medskip\noindent{\em Timelike $W_\mu$}

\medskip\noindent
For timelike $W_\mu$ we may always arrange for its only nonzero component to lie in the time direction: $W_\mu = W \, \delta_\mu^t$. This ensures that the unbroken $SO(2)$ symmetry is simply rotations in the $x-y$ plane. In this case the potential for $S$ would have the form
\bea
 V &=& \frac{W_\mu x^\mu}\ell = \frac{W t}{\ell} = W \, \cosh \tau = W \left( \frac{1 + \tanh^2 \frac{\tau}{2} }{1 - \tanh^2 \frac{\tau}{2}} \right) \nn\\
 &=& W \left( \frac{ 1 + |z|^2}{1 - |z|^2} \right) = \frac{W(1 + 4 S^* S)}{2  (S + S^*)}\,.
\eea
Unlike the situation for the sphere, this potential diverges as $|S| \to 0$ or $\infty$, corresponding to the divergence of the initial potential as $|z| \to 1$ or $\tau \to \infty$. Because it diverges in these limits this provides an example where the entire framework of approximate symmetry also begins to break down there too. (Considering a spacelike $W_\mu$ yields a similarly pathological potential.)

An example of a purely bounded potential comes from the inverse of the one just considered:
\bea
 V &=& \frac{W^2\ell}{W_\mu x^\mu} = -\frac{W\ell}{t} = -W \, \sech \tau = W \left( \frac{1 - \tanh^2 \frac{\tau}{2}}{1 + \tanh^2 \frac{\tau}{2} } \right) \nn\\
 &=& -W \left( \frac{1 - |z|^2}{ 1 + |z|^2} \right) = -\frac{2 W (S + S^*)}{1 + 4 S^* S}\,.
\eea

\medskip\noindent{\em Null $W_\mu$}

\medskip\noindent
For null $W_\mu$ we may always choose the spatial components to point in the $x$-direction: $W_t = -W_x = W$. In this case the potential would have the form
\be
 V = \frac{W_\mu x^\mu}\ell = \frac{W (t - x)}{\ell} = W \Bigl( \cosh \tau - \sinh \tau \, \cos \theta \Bigr) \,,
\ee
and so switching to projective coordinates using
\be
\cosh\tau = \frac{1+\tanh^2(\tau/2)}{1-\tanh^2(\tau/2)} \,,\quad \sinh\tau = \frac{2\,\tanh(\tau/2)}{1-\tanh^2(\tau/2)} \,,\quad \cos\theta = \frac{e^{i\theta}+e^{-i\theta}}2
\ee
gives
\be
V = W \, \frac{(1-z)(1- z^*)}{1-z z^*} = \frac{W}{S+ S^*} \,.
\ee
Thus we find a potential in this case that remains bounded for large $|S|$, and for large fields yields an exponential potential when expressed using canonically-normalized scalars.

\section{Supersymmetric $SU(1,1)/U(1)$ models}
\label{app:susy}

This appendix lays out some simple supersymmetric potentials built around the nonlinear realization of $SU(1,1)$. This shows some of the ways that supersymmetry can supplement the protection of pseudo-Goldstone boson potentials. (We rely heavily on the conventions of \cite{SugraText} in what follows.)

We consider the following K\"ahler potential:
\be
K = -p \log(S+S^*) \,.
\ee
The three isometries of the corresponding K\"ahler metric $g_{S S^*} := \pd_\ssS \pd_{\ssS^*} K$ are captured by the following holomorphic Killing vectors:
\be
 k_1^\ssS = i \,,\quad k_2^\ssS = S \,,\quad k_2^\ssS = iS^2 \,.
\ee
An isometry of the K\"ahler metric is one for which $\delta_\ssA K = \theta^\ssA [ r_\ssA(S) + r^*_\ssA( S^*)]$; hence using
\bea
 \delta_1 K &=& \theta^1 \big(k_1^\ssS \pd_\ssS + k_1^{\ssS^*} \pd_{\ssS^*} \big) K = 0 \\
 \delta_2 K &=& \theta^2 \big(k_2^\ssS \pd_\ssS + k_2^{^*\ssS} \pd_{\ssS^*} \big) K = -p\theta^2 \\
 \delta_3 K &=& \theta^3 \big(k_3^\ssS \pd_\ssS + k_3^{\ssS^*} \pd_{\ssS^*} \big) K = -ip\theta^3 (S-S^*) \,,
\eea
we can read off
\be \label{rAs}
 r_1(S) = -i\xi_1 \,,\quad r_2(S) = -p/2-i\xi_2 \,,\quad{\rm and}\quad r_3(S) = -ip\,S-i\xi_3 \,,
\ee
respectively, which are unique up to the real constants $\xi_\ssA$. Writing $k_\ssA := k_\ssA^\ssS \pd_S$, one can also verify that the Killing vectors satisfy the $SU(1,1)$ algebra:
\be
 [k_1,k_2] = k_1 \,,\quad [k_2,k_3] = k_3 \,,\quad [k_3,k_1] = 2 \, k_2 \,.
\ee

\subsection*{$D$-term Potential}

Assuming (some of) these symmetries are gauged in a supersymmetric way using the gauge fields $A_\mu^\ssA$ with kinetic term
\be
 \cL_{k\ssA} = -\frac14 {\rm Re} f(S) \, F_{\mu\nu}^\ssA F^{\mu\nu}_\ssA \,,
\ee
the $D$-term potential becomes
\be
 V_\ssD = \frac1{2 \, {\rm Re} f(S)} \,\cP_\ssA \cP^\ssA
\ee
where the moment maps $\cP_\ssA$ are given by
\be
 \cP_\ssA = i \big(k_\ssA^\ssS \pd_\ssS K - r_\ssA\big) \,.
\ee
Substituting, we find
\be
 \cP_1 = \frac p{S+S^*} - \xi_1 \,,\quad \cP_2 = \frac p{2i} \left(\frac{S-S^*}{S+S^*}\right) - \xi_2 \,,\quad \cP_3 = -p \left(\frac{SS^*}{S+S^*}\right) -\xi_3 \,.
\ee

We impose the `equivariance relation'
\be
 \big(k_\ssA^\ssS \pd_\ssS + k_\ssA^{\ssS^*} \pd_{\ssS^*} \big) \cP_\ssB = {f_{\ssA\ssB}}^\ssC \cP_\ssC
\ee
to fix the $\xi_\ssA$'s. (This ensures that the $\cP_\ssA$ transform in the adjoint representation of the group, and is necessary for a supersymmetric coupling to the gauge fields' $D$-terms.) This gives $\xi_\ssA = 0$ for all $A$.
Therefore, the moment maps take the following form:
\be
\cP_1 = \frac p{S+S^*} \,,\quad \cP_2 = \frac p{2i} \left(\frac{S-S^*}{S+S^*}\right) \,,\quad \cP_3 = -p \left(\frac{SS^*}{S+S^*}\right)  \,.
\ee
(The absence of $\xi$'s due to the above is consistent with the fact that---although $SU(1,1)$ has an abelian subgroup---the coset group $SU(1,1)/U(1)$ which $S$ describes does not.)

Therefore, if we take $f(S) = S$, and let $S := s+ia$ where $s$, $a$ are real fields, then we find that the possible $D$-term potentials are given by
\be
V_1(s) = \frac {p^2}{8s^3} \,,\quad V_2(s) = \frac{p^2a^2}{8 s^3} \,,\quad V_3(s) = \frac{p^2(s^2+a^2)^2}{8s^3} \,.
\ee
Any of these can be selected by deciding which isometries we choose to gauge.

Avoiding any dependence on $a$ in the potential, we find that the inflaton lagrangian in the case where we gauge only $k_1$ is
\be
 \cL_s = -\frac p{4s^2} \, \pd_\mu s \pd^\mu s - V_1(s) = -\frac12 \, \pd_\mu \phi \pd^\mu \phi - \frac{p^2}8 \, e^{-3\alpha \phi}
\ee
where, past the second equality, we introduce
\be
 \phi:=\frac1\alpha \, \ln s \qquad\leftrightarrow\qquad s := e^{\alpha\phi} \qquad{\rm with}\qquad \alpha = \sqrt{\frac p2} \,.
\ee
In the above, the axion is eaten by the gauge field $A_\mu^1$. Requiring $3\alpha = 0.1$ gives
\be
 p = 1/450 \simeq 0.002 \,.
\ee

\subsection*{$F$-term Potential}

Any superpotential must transform under gauge transformations such that the quantity $e^K \, W \ol W$ is gauge invariant:
\be
\delta_\theta W = \theta^\ssA k_\ssA^\ssS \pd_\ssS W = -\theta^\ssA r_\ssA W \,.
\ee
For each of the isometries, the following conditions then hold:
\begin{align}
(A=1):\quad& i\pd_\ssS W = 0 \quad\implies\quad W = W_0 := {\rm const.} \\
(A=2):\quad& S \pd_\ssS W = \frac p2 \,W \quad\implies\quad W = W_0 \, S^{p/2} \\
(A=3):\quad& iS^2 \pd_\ssS W = ip S W \quad\implies\quad W = W_0 \, S^p \,.
\end{align}
Focussing on the case where $A=1$, we see that the $F$-term potential must take the form
\be
V_\ssF = |W_0|^2 e^K \big(g^{SS^*} \pd_S K \pd_{S^*} K -3 \big) = (p-3) \frac{|W_0|^2}{(S+S^*)^p}
\ee

\end{document}